\documentclass[reprint,12pt,NumberedRefs]{JASAnew}

\usepackage{siunitx}
\usepackage{epstopdf}

\begin{document}

\title[Gas-filled rotating spheroid acoustic spectra]{Acoustic spectra of a gas-filled rotating spheroid}
\author{Sylvie Su}
\author{David C\'ebron}
\email{david.cebron@univ-grenoble-alpes.fr} 
\author{Henri-Claude Nataf}
\author{Philippe Cardin}
\affiliation{Univ. Grenoble Alpes, Univ. Savoie Mont Blanc, CNRS, IRD, IFSTTAR, ISTerre, 38000, Grenoble, France}
\author{J\'er\'emie Vidal}
\affiliation{Department of Applied Mathematics, School of Mathematics, University of Leeds, Leeds, LS2 9JT, UK}
\author{Max Solazzo}
\author{Yann Do}
\affiliation{Univ. Grenoble Alpes, Univ. Savoie Mont Blanc, CNRS, IRD, IFSTTAR, ISTerre, 38000, Grenoble, France}

\date{\today}

\begin{abstract}
The acoustic spectrum of a gas-filled resonating cavity can be used to indirectly probe its internal velocity field. This unconventional velocimetry method is particularly interesting for opaque fluid or rapidly rotating flows, which cannot be imaged with standard methods. This requires to (i) identify a large enough number of acoustic modes, (ii) accurately measure their frequencies, and (iii) compare with theoretical synthetic spectra. Relying on a dedicated experiment, an air-filled rotating spheroid of moderate ellipticity, our study addresses these three challenges. To do so, we use a comprehensive theoretical framework, together with finite-element calculations, and consider symmetry arguments. We show that the effects of the Coriolis force can be successfully retrieved through our acoustic measurements, providing the first experimental measurements of the rotational splitting (or Ledoux) coefficients for a large collection of modes. Our results pave the way for the modal acoustic velocimetry to be a robust, versatile, and non-intrusive method for mapping large-scale flows.
\end{abstract}

\keywords{acoustic resonances, spheroid, velocimetry, rotation}

\maketitle

\section*{\label{sec:Intro} Introduction}
The acoustic response of a resonant cavity is directly related to its geometry and physical properties. 
In the past, seismology has used those free oscillations to probe the Earth's interior \citep{backus1961rotational,pekeris1961rotational}.
This method has successfully contributed to determine properties such as density, elasticity and anisotropy parameters taking into account the rotation and the shape of the Earth \citep{dahlen1968normal}. 
Similarly, eigenmodes measurements have also been developed in helio \citep{thompson1996differential} and asteroseismology \citep{ledoux1951nonradial}, giving insights on the stellar interiors \citep{christensen2002helioseismology,aerts2010asteroseismology}.

Inspired by asteroseismology, a recently presented laboratory velocimetry method relies on acoustic resonances to invert flows within a spherical shell \citep{triana2014}. By contrast with usual velocimetry techniques, for example particle tracking methods such as Particle Image Velocimetry (PIV) \citep{wereley2010recent} or methods based on Doppler effect such as laser Doppler \cite{albrecht2013laser} and ultrasound Doppler velocimetry \citep{brito01,tigrine2018torsional}, modal acoustic velocimetry does not require a seeded fluid as it measures the fluid acoustic response directly \citep{triana2014}. This is of great interest for all experiments where seeding the fluid is not practically feasible,  in particular for gases \citep{melling1997tracer}.
Another benefit is that modal acoustic velocimetry is a non-intrusive technique providing flow fields in the whole volume, even in opaque fluids. 

Interested in zonal jets, their formation, geometry, and amplitude in geo-astrophysical flows \citep{beebe1994characteristic, manneville1996,cabanes2017laboratory}, we built a reduced model that obeys similar force balances. Our apparatus is a rotating oblate spheroid called  ZoRo (Zonal flows in Rotating fluids). 
It has been shown that in some cases \citep{busse1994}, the zonal flows are geostrophic, \textit{i.e.} invariant along the axis of rotation of the fluid, forming concentric cylindrical shells \cite{kaspi2018jupiter,kong2018origin}. Such flows are expected to have significant influence on acoustic global modes \cite{aerts2010asteroseismology,triana2014}.

Following previous experimental studies \citep{Ecotiere_Tahani_Bruneau_2004, triana2014}, we stress that three challenges must be met in order to effectively map these flows: (i) identify a large enough number of modes; (ii) accurately measure their frequencies; (iii) compare with theoretical synthetic spectra that take into account the effect of rotation. The present study addresses these three goals, and focuses on measuring and computing the effect of rotation on a large collection of acoustic modes in a spheroid.

The paper is organised as follows. 
Section~\ref{sec:Exp} presents the ZoRo experimental apparatus. 
In Section~\ref{sec:Theory}, we present the theoretical framework we use to predict the acoustic response of our experiment. 
In Section~\ref{sec:Expspectra}, we analyse and interpret experimental acoustic spectra. 
To refine this interpretation, in Section~\ref{sec:Secorder} we use a second-order geometrical perturbation theory, and test it against spectral and finite-element calculations.
Then, in Section~\ref{sec:Rot} we introduce rotation within the experiment, and compare the measured acoustic response to perturbation theory.
Finally, Section~\ref{sec:Conclu} summaries the different approaches and presents some perspectives. 

\section{The ZoRo experiment}\label{sec:Exp}
\begin{figure}[ht]
\includegraphics[width=0.35\textwidth]{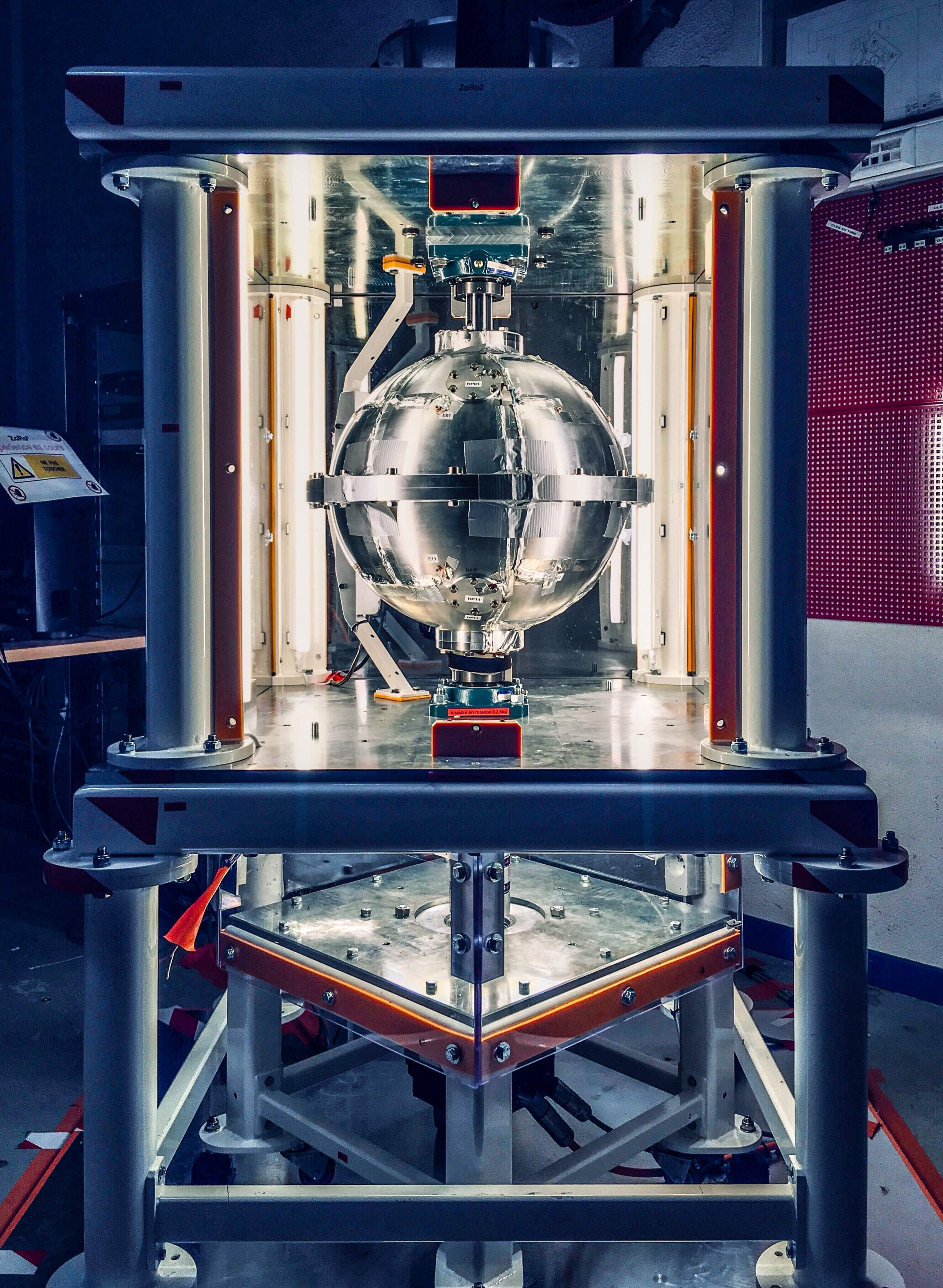} \\
\vspace*{1.0mm}
\includegraphics[width=0.35\textwidth]{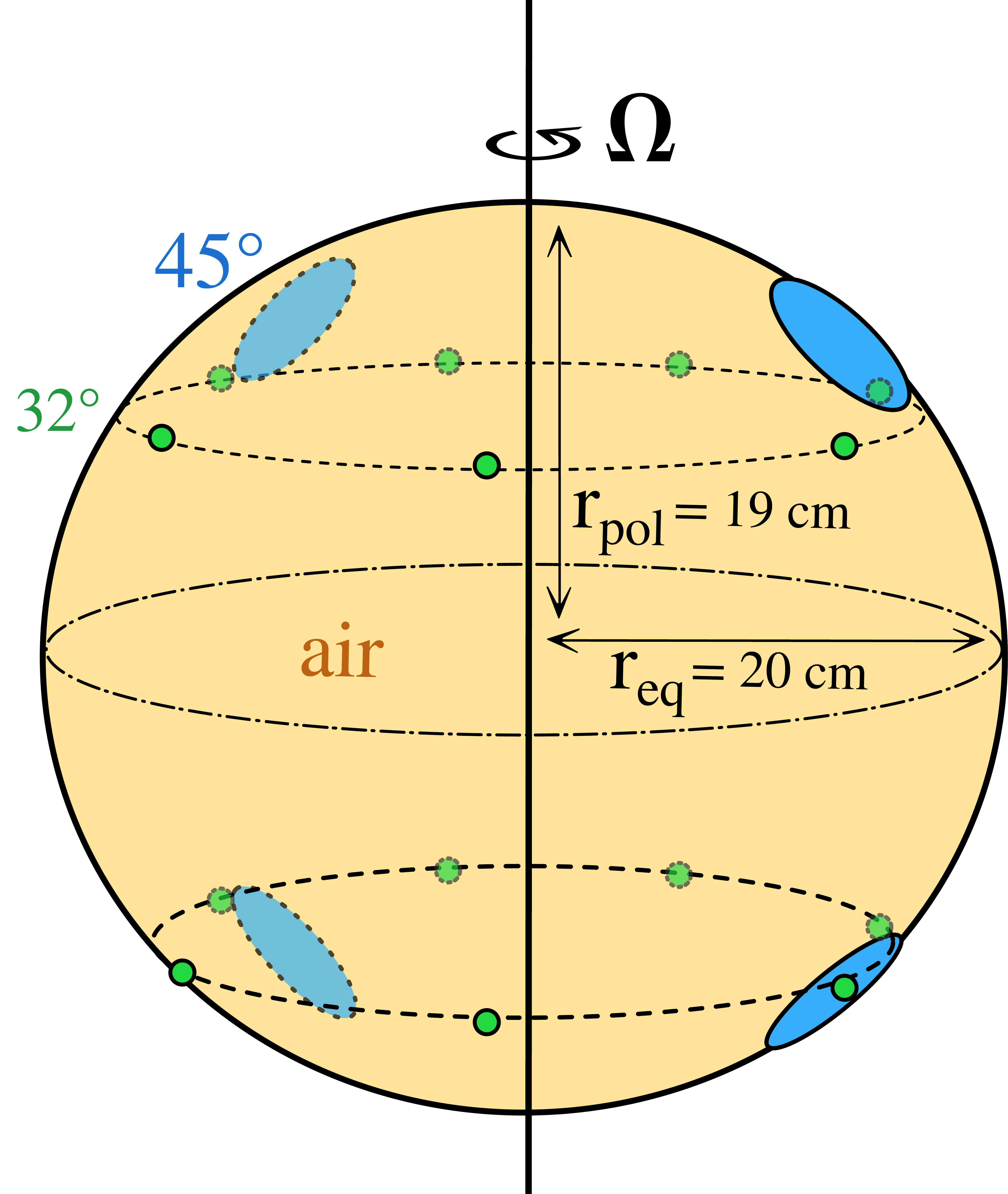}
\caption{\label{fig:apparatus}{Photo (top) and sketch (bottom) of the ZoRo experimental apparatus. Four speakers (blue) are at $\pm 45^\circ$ latitude. The acoustic response of the gas-filled resonator (yellow) is measured with electrets (green), at $\pm 32^\circ$  (colour online). The instrumentation is installed on the rotating apparatus, which can rotate around its symmetry axis up to $|\boldsymbol{\Omega}|/(2\pi) =f_{\Omega}=30$~Hz.}}
\end{figure}

The experimental apparatus ZoRo is a 1~cm thick shell made with aluminium based alloy (Thyssenkrupp by Constellium cast, MecAlu+ 7000 series) enclosing an axisymmetric oblate spheroidal cavity of equatorial radius $r_{eq} = 20$~cm and polar radius $r_{pol} = 19$~cm filled with air (see Figure \ref{fig:apparatus}). The ellipticity is defined as $e = (r_{eq}- r_{pol})/r_{eq} = 0.05$. 

Special care has been given during the fabrication process, especially to the junction plane between the two hemispheres and the coaxiality (both between the hemispheres and the shafts) which have been identified in the literature as common defaults \cite{mehl1986acoustic}. This allows us to ensure the dimensions of the shell down to 0.1~mm.
Chosen excerpts of technical drawings are provided in Supplementary material \cite{supmatdraw}.

The spheroid's revolution axis is mounted on the shaft of an electric motor (ref. Kollmorgen AKM73Q) through a vibration-reducing jaw-type coupler (ref. ROTEX SH38 from KTR), allowing rotation rates up to $50$~Hz, or 3000~revolution per minute (rpm).
Static balancing and dynamic balancing with the shell rotating up to $70$~Hz in the experimental conditions, reveal less than 0.2~g of unbalanced mass (for a total rotating spheroid mass of $32.5$~kg). 

Acoustic pressure is measured by electret microphones (ref. Projects Unlimited \hbox{TOM-1545P-R}) connected to a mixing table (ref. TASCAM US-16x08). Acoustic waves are produced by 36~mm-diameter audio speakers with working range 400-6000~Hz (ref. Multicomp MCKP3648SP1-4758) connected to a sound card (ref. Asus TeK, Xonar DGX), through 20~W amplifiers (LEPY LP-808). 
Acoustic resonances are excited by chirps sweeping over the frequencies of interest. At $T=20^\circ$C and atmospheric pressure, the cavity's fundamental mode is at $578$~Hz.
Both source and data files are sampled at 44.1~kHz and written in 16-bits in the uncompressed audio specific Waveform Audio File Format (WAVE), corresponding to standard CD audio quality.

The instrumentation is in contact with the gas in through holes in the aluminium shell. Air-tightness is ensured by custom made plastic joints on each hole and at the equatorial seam of the shell.
Additional holes (see ), provided for future temperature and pressure sensors, are plugged closed.
The apparatus can accommodate up to four speakers and fourteen electrets, half on each hemisphere, at respectively $\pm45^{\circ}$ and $\pm32^{\circ}$ latitude (Fig.~\ref{fig:apparatus}).
All speakers are on the same meridional plane, electrets are evenly spaced on their latitude; lower and upper hemispheres are strictly symmetric (for technical drawings, see Supplementary material\cite{supmatdraw}).
Electric signals are transmitted from the rotating to the laboratory immobile frames through two slip rings (PSR-HSC-36 and PSRT-38H-24 from Panlink) with gold-gold contacts. 

\section{\label{sec:Theory}Theoretical framework}
\subsection{Governing equations and boundary conditions} \label{sec:govEq}
We consider the small perturbations $[\rho,p,\boldsymbol{u},T]$ upon a homogeneous basic (background) state of density $\rho_0$, pressure $p_0$, velocity $\boldsymbol{u}_0$, and temperature $T_0$.
We thus linearize the governing flow equations for a Newtonian fluid (see Supplementary material \cite{supmatmain}),
assuming $\boldsymbol{u}_0=\boldsymbol{0}$ in the reference frame that is rotating at the constant rate $\boldsymbol{\Omega}= \Omega \, \boldsymbol{1}_z$ (with $\boldsymbol{1}_z$ the unit vector along the polar axis).
For uniform viscous and thermal diffusivities, the equations are \citep{blackstock2001fundamentals}
\begin{subequations}
\allowdisplaybreaks
\begin{eqnarray}
\frac{\partial \rho}{\partial t} &=& -\nabla \cdot (\rho_0 \boldsymbol{u}) , \label{eq:1} \\
\rho_0 \frac{\partial \boldsymbol{u} }{\partial t}  & =& - 2 \rho_0 \, \boldsymbol{\Omega} \times \boldsymbol{u} 
-\nabla p + \mu \nabla^2 \boldsymbol{u} \nonumber \\& & \quad  + \left(\mu_B+\frac{\mu}{3} \right) \nabla (\nabla \cdot \boldsymbol{u})  , \label{eq:2}\\
\rho_0 C_p  \frac{\partial T}{\partial t}    
&=&  \alpha T_0 \frac{\partial p}{\partial t}  + \lambda \nabla^2 T  , \label{eq:3} \\
\rho&=&\rho_0(\beta_T p -\alpha T) , \label{eq:4} 
\end{eqnarray}
\end{subequations}
using the usual underlying assumptions (which removes $p_0$ from the dynamical equations, see the Supplementary material \cite{supmatmain} for details).
In equations (\ref{eq:1})-(\ref{eq:4}), $t$ is the time, $\lambda$ is the thermal conductivity, $\alpha$ is the isobaric coefficient of thermal expansion, $C_p$ is the specific heat at constant pressure, $\beta_T=\gamma \beta_s$ is the isothermal compressibility, where $\gamma$ is the specific-heat ratio and $\beta_s$ is the isentropic compressibility, $\mu$ is the dynamic (shear) viscosity, and with the bulk viscosity $\mu_B$, which is related to the second coefficient of viscosity $\mu_B-2 \mu /3$. Note that the sound speed $c$ is then given by $c=(\rho_0 \beta_s)^{-1/2}$.

To solve the system of equations (\ref{eq:1})-(\ref{eq:4}) we supplement them with boundary conditions. Assuming that the sound speed is much larger in the container than in the cavity, we consider a rigid spheroidal boundary and impose the no-slip boundary condition $\boldsymbol{u}=\boldsymbol{0}$. We also assume that the container thermal conductivity is much larger than $\lambda$ and impose an isothermal boundary condition.

Finally equations (\ref{eq:1})-(\ref{eq:4}) can be solved in the frequency domain by looking for time periodic solutions for $[\rho,p,\boldsymbol{u},T]$. With an eigensolver, we can then obtain the fluid eigenmodes such as the acoustic modes \cite[lower frequency inertial modes are also obtained in rotating fluids, see e.g.][]{vidal2019acoustics}. One can also rather calculate the spectral fluid response if we define excitation sources for $[\rho,p,\boldsymbol{u},T]$, e.g. by modelling the experimental speakers using (imposed) time periodic flows at the speaker locations. 

\subsection{\label{subsec:principles}Principles of perturbation theory}
We seek the fluid normal eigenmodes with a harmonic time dependence $\exp(\mathrm{i} \omega t)$, where $\omega$ is the (possibly complex) pulsation. The system of governing equations (\ref{eq:1})-(\ref{eq:4}) can be cast into a symbolic eigenvalue equation as
\begin{equation}
\label{eq:H}
\mathcal{H} \boldsymbol{v} = -\omega^2 \boldsymbol{v},
\end{equation}
where $\mathcal{H}$ is a complex vectorial operator linear in $\boldsymbol{v}$, which can be the velocity $\boldsymbol{u}$ or the Lagrangian displacement for isentropic fluids \citep[e.g.][]{ledoux1951nonradial}; this equation being supplemented by appropriate boundary conditions.
Analytical solutions of the complete problem are usually not available.
A convenient and classical approach is to define a reference model, solution of equation (\ref{eq:H}) with $\mathcal{H}=\mathcal{H}_0$ and simple boundary conditions,
for which analytical solutions are available. The missing terms of the original operator $\mathcal{H}$ 
and boundary conditions are then treated as perturbations\cite{morse1953methods,Dahlen_1998}.

After choosing our reference model, 
we successively consider perturbations due to (i) Coriolis force, (ii) elliptical shape of the rigid shell, and (iii) dissipation.
First-order perturbations are often sufficient, with the advantage that the frequency perturbation can be computed without calculating the corresponding eigenvector.
However, accurate calculation of the modes may require higher-order perturbations \citep{vidal2019acoustics}, for instance for the Coriolis force \citep{backus1961rotational} or the elliptical shape \cite{mehl2007acoustic}.

\subsection{\label{subsec:refstate} Modes of the reference model}
We aim to study theoretically the homogeneous motionless basic state, and its eigenmodes, defined in section \ref{sec:govEq}. 
Without loss of generality, we tackle this problem by considering, for our reference model, a diffusionless gas enclosed in a non-rotating sphere of (arbitrary) radius $a$. 
The latter value remains unspecified for the moment (see below). 
Equation (\ref{eq:H}) becomes $c^2\, \nabla \left( \nabla \cdot \boldsymbol{u}\right) = -\omega^2 \boldsymbol{u}$.

We introduce the velocity potential $\Psi$ such that $\boldsymbol{u} = \nabla\Psi$, which relates to the acoustic pressure in the gas by $p=-\partial \Psi/\partial t$. Then, problem (\ref{eq:H}) reduces to the Helmholtz equation

\begin{equation}
c^2\nabla^2 p = -\omega^2 p.
\label{eq:pwave_k}
\end{equation}
In spherical coordinates $(r,\theta,\varphi)$, 
the solutions of equation (\ref{eq:pwave_k}) that satisfies the boundary condition $\partial p/\partial r=0$ are
\begin{equation}
p(r,\theta,\varphi) = p_0 \, {j_l}(k_{nl}r) \, \mathcal{\mathcal{Y}}_l^m(\theta,\varphi),
\label{eq:sphericalsol}
\end{equation}
where $k_{nl}$ is the radial wavenumber,
$\mathcal{\mathcal{Y}}_l^m$ is the spherical harmonic of degree $l$ and order $m$ and yields the angular dependency, ${j_l}$ is the spherical Bessel function of the first kind of degree $l$ and carries the radial dependency (we remind that $|m| \leq l$). The boundary condition $\left.\mathrm{d}j_l(k_{nl}r)/\mathrm{d}r\right|_{r=a} = 0$ completes the quantification of the modes, yielding a radial mode number $n$, labelling the $n^{th}$ zero of $\mathrm{d}j_l(k_{nl}r)/\mathrm{d}r$.

One $(n,l,m)$ triplet fully characterises a given acoustic mode of our spherical enclosure.
We denote it ${_n}\mathcal{S}_l^m$, following the convention of seismologists \citep{Dahlen_1998}.
The spherical symmetry of our reference model implies that the frequencies $f_{nl}$ of the modes are independent of the azimuthal mode number $m$.
One gets $f_{nl} = c \, k_{nl}/(2\pi)$.
The perturbations we consider lift this degeneracy partially or totally.
They also modify the corresponding eigenvectors.
However, since we consider rotation of the spheroid around its symmetry axis, we can still label the perturbed modes as ${_n}\mathcal{S}_l^m$.

\subsection{\label{subsec:perturb} Perturbations}
\subsubsection{\label{par:coriolis} Coriolis splitting}
The Coriolis term $- 2 \rho_0 \, \boldsymbol{\Omega} \times \boldsymbol{u}$ in equation (\ref{eq:2}) breaks the $\pm\varphi$-symmetry.
As a consequence, the degenerate spectral peak of a ${_n}\mathcal{S}_l$ multiplet splits into $2l+1$ peaks, corresponding to all ${_n}\mathcal{S}_l^m$ singlets.
According to first-order perturbation theory, the frequency shift $\delta_\Omega$ of a ${_n}\mathcal{S}_l^m$ singlet with respect to the ${_n}\mathcal{S}_l$ degenerated frequency is \cite{ledoux1951nonradial}
\begin{equation}
    \delta_{\Omega} = - m \, f_\Omega \, C_{nl},
    \label{eq:deltaomROT}
\end{equation}
where $0\leq C_{nl}<1$ are called the Ledoux coefficients, and noting $f_\Omega=\Omega/(2 \pi)$.
Their expression is 
\citep[see equation 3.361 in][]{aerts2010asteroseismology} 
\begin{equation}
C_{nl} = \frac{\int_{0}^{a} \left[ 2 \, \xi_r \, \xi_h + \xi_h^2\right] r^2 \mathrm{d}r}{\int _{0}^{a} \left[ \xi_r^2 + l(l+1) \, \xi_h^2\right] r^2 \mathrm{d}r},
\label{eq:Ledoux}
\end{equation}
with $\xi_r = \mathrm{d}j_l(k_{nl}r)/\mathrm{d}r$ and $\xi_h = j_l(k_{nl} r)/r$.
We list the values of Ledoux coefficients for a chosen collection of acoustic modes in Supplementary material \cite{supmatmain}.

One of the main goals of our study is to retrieve the frequency splittings due to the Coriolis force in the ZoRo experiment, and compare them to predictions \citep{ledoux1951nonradial}.

Note that the splitting of acoustic modes in a gas-filled rotating cylinder has been used to design gyrometers \cite{Ecotiere_Tahani_Bruneau_2004}, and that seismologists \cite{backus1961rotational} developed the second-order perturbation theory for the Coriolis term, in order to match the observed splitting of the lowest normal modes of the Earth.

\subsubsection{\label{par:geom} Geometrical splitting}
Coriolis splitting is very small for acoustic overtones with large $n$.
In order to measure the splitting of an individual ${_n}\mathcal{S}_l^{\pm m}$ doublet, it is desirable to separate it from the other doublets of its ${_n}\mathcal{S}_l$ multiplet.
This was the reason for designing the ZoRo experiment as an axisymmetric oblate spheroid.

First-order perturbation theory for elliptical flattening show that the induced frequency splitting $\delta_{geom}$ is quadratic in $m$.
Introducing $\gamma_{nl}$ the ellipticity splitting coefficient, the frequency perturbation writes \cite{dahlen1968normal,Dahlen_1976}
\begin{equation}
    \delta_{geom} = f_{nl} \left(-\frac{1}{3}l(l+1) + m^2 \right) \gamma_{nl} \, e,
\end{equation}
where $f_{nl}$ is the frequency of the ${_n}\mathcal{S}_l$ mode in the spherical reference model (with the same volume as the spheroid), and $e$ the ellipticity.
We give the expression of $\gamma_{nl}$, and its value for a collection of modes, in Supplementary material \cite{supmatmain}.

We obtain the same first-order results using formulas from either Dahlen \cite{Dahlen_1976} or Mehl \cite{mehl2007acoustic} provided the radius of the reference sphere is the same.
We will see later that we need to consider second-order geometrical perturbations, which have been derived by Mehl \cite{mehl2007acoustic} using Morse and Feshbach's formalism \cite{morse1953methods}.

\subsubsection{\label{par:diffusion} Dissipation}
In order to construct realistic spectra that can be compared with observations, it is essential to take into account dissipative effects, which control the width and the amplitude of the resonance peaks \citep{moldover1986gas, trusler1991physical}.
Dissipation of acoustic modes in air is dominated by viscous friction and heat diffusion at solid boundaries, with minor contributions from bulk viscosity \cite{moldover1986gas}.

Considering the spherical reference model, perturbation methods \cite{Kirchhoff_1868, Jordan_2016} 
provide estimates for the diffusion effect $g_{nl} = g_{bound}+g_{bulk}$ , where $g_{bound}$ and $g_{bulk}$ are respectively the boundary and bulk contributions. The mode complex frequencies can then be written as $f_{nl}-g_{bound} + \mathrm{i} g_{nl} $, where $g_{bound}$ and $g_{bulk}$ take positive (real) values \cite{moldover1986gas}. Note that the imaginary part (i.e. the damping of the mode) is due to $g_{nl}$  but only $g_{bound}$ affects the real part, reducing the (real) eigenfrequency. Assuming that the container is a much better heat conductor than the gas, dissipation depends only upon physical properties of the gas: kinematic viscosity $\nu=\mu/\rho$ and bulk viscosity $\nu_{bulk}=\mu_B/\rho_0$, thermal diffusivity $\kappa=\lambda/(\rho_0 C_p)$, and the adiabatic index $\gamma$. The $g_{bound}$ contribution is then given in Moldover \cite{moldover1986gas}, equation (42), by 
\begin{equation}
g_{bound} =  \frac{f_{nl}}{2 a} \frac{(\gamma - 1) \, d_T + l(l+1) \, d_U / z_{nl}^2}{1 - l(l+1) / z_{nl}^2},
\end{equation}
which, in our case, is much larger than the bulk contribution, corrected by \cite{guianvarc2009acoustic}
\begin{equation}
g_{bulk} = \frac{z_{nl}^2}{4 \pi a^2} \left[ (\gamma - 1) \kappa +\frac{4}{3} \nu + \nu_{bulk} \right],
\label{eq:bulk}
\end{equation}
where  $z_{nl}=k_{nl} a$ is the dimensionless radial wave number, and $d_T = \sqrt{\kappa/(\pi f_{nl})}$ and $d_U = \sqrt{\nu/(\pi f_{nl})}$ are the thicknesses of the thermal and viscous boundary layers respectively. 
We use air thermodynamic properties \citep{Cramer_2012}, see also \url{https://encyclopedia.airliquide.com/fr/air}, 
to calculate these contributions and list our values of $g_{nl}$ coefficients for a collection of acoustic modes in the Supplementary material \cite{supmatmain}.

\subsubsection{\label{par:other} Other perturbations}
Other effects might influence the spheroid's acoustic spectrum. We list and explore a few of them that are mentioned in the literature as potentially relevant.
They include uneven shell surface finishing, presence of holes in the shell or a seam between the two hemispheres. All of those are expected to be negligible for our apparatus \cite{moldover1986gas}. 

The finite elasticity of the container may modify significantly the acoustic resonances when acoustic and elastic eigenfrequencies are close \citep{moldover1986gas}. 
Influences of the shell elasticity of our apparatus have been estimated with analytical calculations \citep{rand1967vibrations,lonzaga2011suppression} and cross-checked with finite-element simulations, as detailed in Supplementary material \cite{supmatmain}.
Both methods predict frequency shift of $0.1-1$~Hz far from the elastic resonances, and much larger when elastic and acoustic resonances are close. However, due to the high complexity of our apparatus (complex shell geometry, presence of screws, mount on an outer frame), accurate predictions of the apparatus elastic eigenfrequencies remain out of reach. Note that, because of its finite elasticity, the container oscillates, leading to sound scattering in the surrounding air. Based on our elastic calculations, we have verified that this sound scattering can also be neglected \cite{supmatmain}.

\subsection{\label{subsec:synthetic} Synthetic spectra}

For air enclosed in a rigid container, effects influencing the complex frequencies can be linearly superposed\citep{moldover1986gas}. Adding the different perturbations to $f_{nl}$, we obtain the predicted  frequency $f_{nlm}$ of each $_n\mathcal{S}_l^m$ mode. Assuming a Lorentzian resonance \cite{moldover1986gas,trusler1991physical}, we compute its contribution $\mathcal{A}_{nlm}(f)$ to the complete frequency spectrum by
\begin{equation}
\mathcal{A}_{nlm}(f) = \left| \frac{A_{nlm}}{(f-f_{nlm}) - \mathrm{i} \, g_{nl}}\right|,
\end{equation}
with
\begin{equation}
A_{nlm} = A  \left|\zeta j_l(k_{nl}a) \cos\left[ m(\varphi_{el}-\varphi_{sp}) \right] \right|,
    \label{eq:pressureamp}
\end{equation}
where $\zeta=\mathcal{P}_l^m\left(\cos\theta_{sp}\right) \mathcal{P}_l^m\left(\cos\theta_{el}\right)$, $A$ is a constant, $\mathcal{P}_l^m$ is the associated Legendre polynomial of degree $l$ and order $m$, $\theta_{sp}$, $\varphi_{sp}$ and $\theta_{el}$, $\varphi_{el}$ the co-latitude and longitude of the speaker (source), resp. electret (receptor), see Supplementary material \cite{supmatdraw} for exact positions of the instrumentation.

\section{\label{sec:Expspectra} Experimental spectra and mode identification}
Using the instrumentation described in Section~\ref{sec:Exp}, we measure the acoustic response of the ZoRo apparatus at rest. Acoustic spectra obtained as such display unambiguous resonances (Fig.~\ref{fig:fullspectrum}) which we aim to interpret.

\subsection{Spectrum interpretation}

\begin{center}
    \begin{figure}[htbp]
\includegraphics[width=\textwidth]{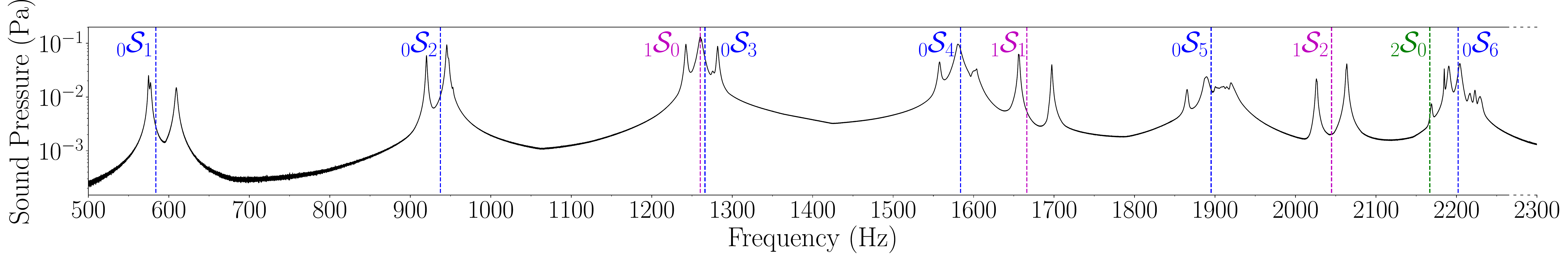}
\includegraphics[width=\textwidth]{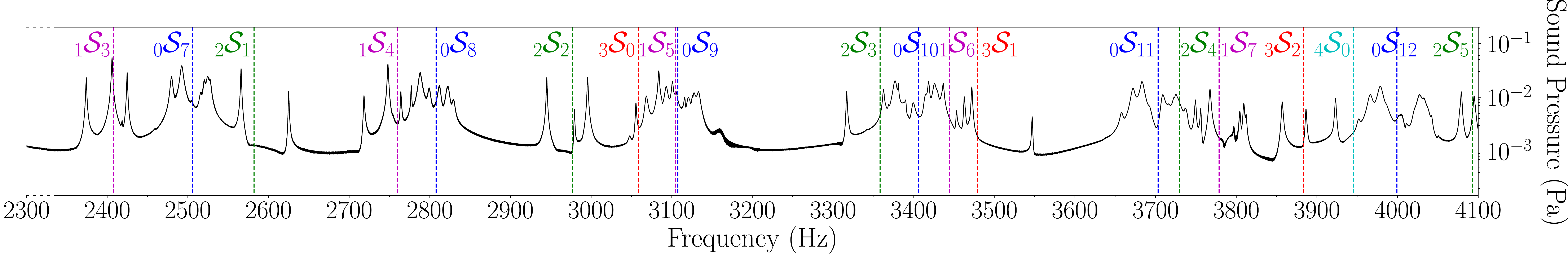}
\caption{\label{fig:fullspectrum}{Experimental acoustic spectrum at rest averaged over all electrets of one hemisphere. The spectrum is continued from the top to the bottom frame. Groups of peaks can be labelled with $_n\mathcal{S}_l$ according to theoretical prediction of a sphere of same volume (dashed lines).}}
\end{figure}
\end{center}

ZoRo's ellipticity is moderate, so at lowest order we expect resonance frequencies to be near those of the sphere of same volume. 
     Eigenfrequencies are proportional to the sound speed $c$, which varies with temperature \citep{zuckerwar1996speed, koulakis2018acoustic}. 
    In order to compare the measured and predicted frequencies, we choose a non-splitting reference peak (${_2}\mathcal{S}_0^0$) and deduce the apparent sound speed, which we then use for the theoretical predictions.
    In the run shown in figure~\ref{fig:fullspectrum}, the reference frequency is $f_{20}=2177$~Hz  and corresponds to $c=346.5$~m.s$^{-1}$ with $a$ taken to be  $a = \sqrt[3]{r_{pol} r_{eq}^2}$ in the following (see Supplementary material \cite{supmatmain}).
    
    In figure~\ref{fig:fullspectrum} we show the experimental spectrum obtained with a continuous linear chirp of duration 90~s with frequencies ranging from 400 to 5000~Hz (which corresponds to the working range of the loudspeakers). 
    We first observe that all peaks are close to the resonances of the reference model. Around the $f_{nl}$, there are groups of peaks, corresponding to degeneracy lifting of the resonances. 
    The lowest frequency modes are clearly separated. We note that those groups contain $l+1$ peaks confirming that the degeneracy is only partially lifted, as predicted by the geometrical perturbation theory \ref{par:geom}.

\onecolumngrid
\twocolumngrid

\subsection{\label{subsec:modeid} Using symmetries to improve mode identification}
\begin{figure}[ht]
\includegraphics[width=0.48\textwidth]{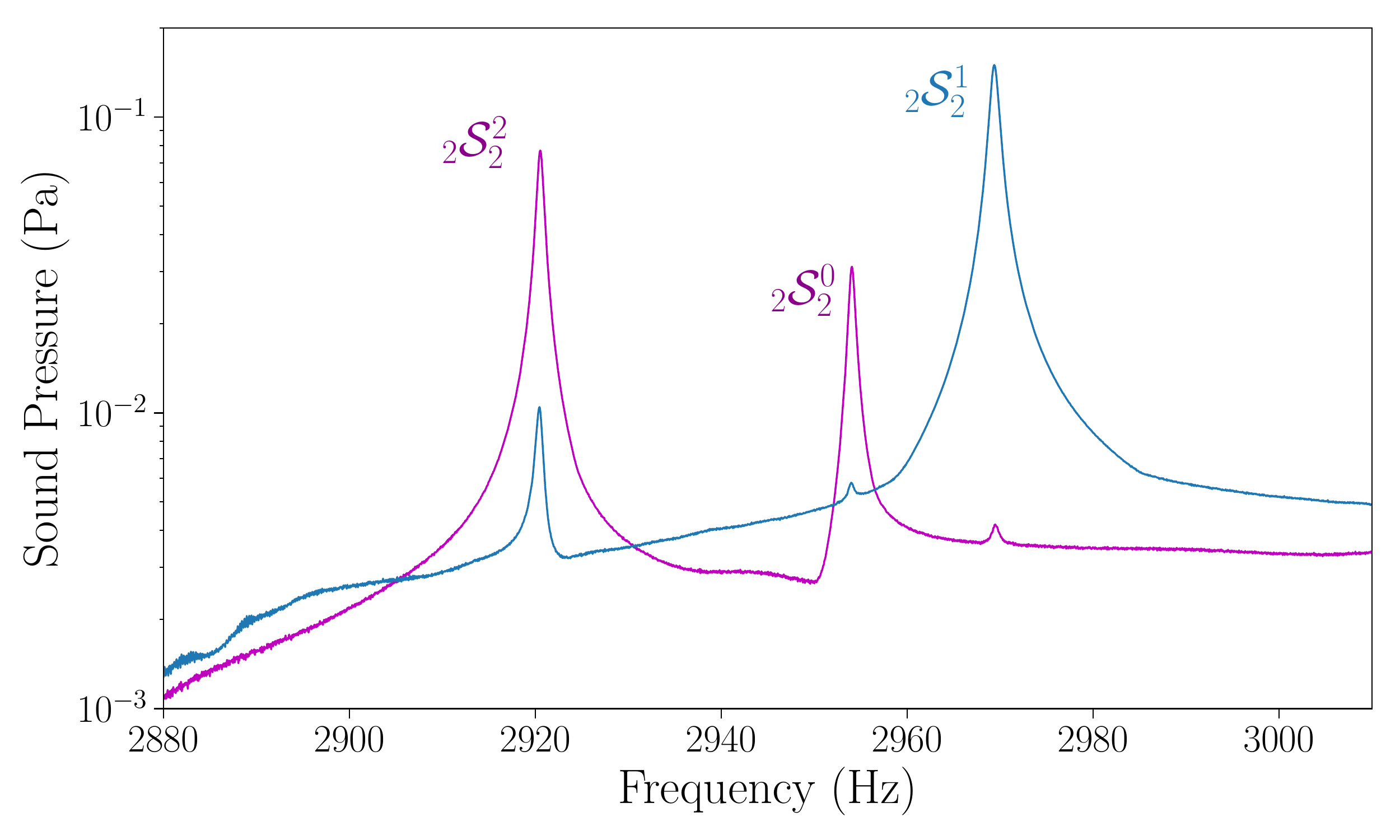}
\caption{\label{fig:2S2invert}{Experimental acoustic spectra centered on the $_2\mathcal{S}_2$ multiplet. Two speakers symmetric with respect to the equator simultaneously play, first in-phase (purple) then in phase opposition (blue). Unexpectedly, $_2\mathcal{S}_2^0$ has lower frequency than $_2\mathcal{S}_2^1$. (colour online)}}
\end{figure}

The symmetric disposition of the instrumentation allows separation of acoustic modes based on their spatial pressure distribution.
Since the pressure field at the resonator's surface is mapped on the spherical harmonics $\mathcal{\mathcal{Y}}_l^m$, as shown in equation (\ref{eq:sphericalsol}), it displays the same symmetries. Acoustic modes with even $l-m$ are symmetric with respect to the equator; those with odd $l-m$ are anti-symmetric \citep{morse1953methods}. 
Acoustic sources impose both wave phase and positions of the anti-nodes, so playing several speakers at the same time strengthens modes with the same symmetries as the source pattern, whereas modes with incompatible symmetries are extinguished, or, in practice, weakened. 
To amplify this effect, the acoustic response of the resonator is systematically measured with at least a pair of equatorially symmetric electrets and we make use of the symmetry properties in the data analysis. Measured signals are summed, or subtracted, to match the source pattern and further attenuate the unwanted modes. 
In figure~\ref{fig:2S2invert}, two speakers symmetric with respect to the equator simultaneously play first in-phase then in antiphase (also called phase opposition). The acoustic response is measured with four pairs of equatorially symmetric electrets. Measured time series of the in-phase speakers are summed for each pair. We compute the frequency spectra which are then averaged, shown in figure~\ref{fig:2S2invert} (purple). We do the equivalent for phase opposition sources signal with subtraction of each pair, shown in figure~\ref{fig:2S2invert} (blue).
The summed spectra of the in-phase source (purple) enhance the even $|m|$ (0 and 2 here), subtracted spectra of phase opposition source (blue) enhance the odd $|m|=1$.

By successively extinguishing even and odd $l-m$ modes, 
we are able to identify, meaning uniquely attribute a $(n,l,m)$ triplet, each peak of the experimental spectrum up to $3500$~Hz (for complete experimental spectrum identification, see Supplementary material \cite{supmatmain}). 

Mode identification allows us to observe several features that seem to be qualitatively consistent throughout the spectrum. For a given $\ _n\mathcal{S}_l$ multiplet, the flattening partly lifts the mode degeneracy and separates the $|m|$ peaks, with higher $|m|$ at lower frequency, which agree with the simple first-order perturbation analysis.
This is consistent with the rule of thumb stating that higher $|m|$ modes are more sensitive to the equatorial region, hence feel a larger radius, yielding a lower frequency.

\section{\label{sec:Secorder} Beyond first-order effects}

Upon more rigorous observations, exceptions to the features mentioned before are visible. We can see on the $_1\mathcal{S}_2$ multiplet, around $2050$~Hz, that the $m=0$ and $|m|=1$ peaks have same frequency, and on $_2\mathcal{S}_2$ around $2950$~Hz, that the $m=0$ peak has lower frequency than $|m|=1$ (Fig.~\ref{fig:2S2invert}).
To explain those deviations, it seems necessary to go beyond first-order effects.
\subsection{\label{subsec:secorder_ellip} Dominant second-order effects: due to the ellipticity}
\begin{figure}[ht]
\includegraphics[width=0.49\textwidth]{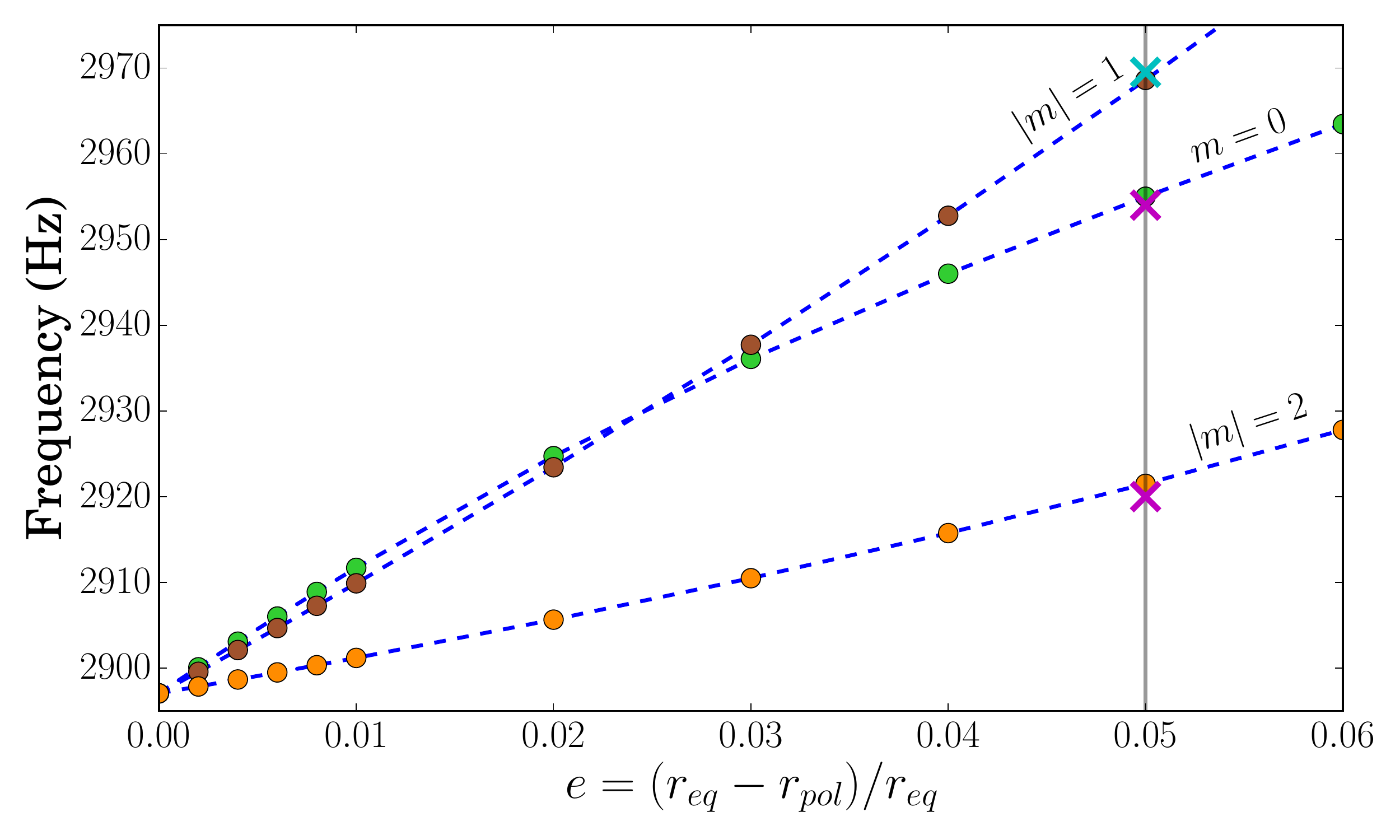}
\caption{Frequency evolution of $_2\mathcal{S}_2^m$ modes with $e$ ($r_{eq}$ being kept constant) given by finite-element calculations (circles) and global polynomial computations \citep{vidal2019acoustics} (dashed lines). Both agree with experimental data (crosses) at the apparatus ellipticity $e= 0.05$ (black line).} 
\label{fig:galcomsolvseps}
\end{figure}

To estimate the dominant second-order effects in the experiment, one can perform order of magnitude calculations. As $e=0.05 \gg \nu/(c r_{eq}) \approx 10^{-7}$ 
the dominant second-order correction is expected to be in $e^2$, originating only from the container geometry second-order perturbation. To check if this effect can explain our observations, we compute diffusionless acoustic resonances of the exact spheroid, as a function of the ellipticity $e$, and we compare them with the experimental observations. 
To do so, we implement two different methods.
First, we use the new global polynomial (Galerkin) approximation of the acoustic modes in rigid ellipsoids \citep{vidal2019acoustics}. Second, we solve the acoustic equation (\ref{eq:pwave_k}) using the built-in acoustic interface of the finite-element commercial software COMSOL Multiphysics.
In figure~\ref{fig:galcomsolvseps}, we plot the frequency evolution as a function of ellipticity for different $m$ of the $_2\mathcal{S}_2$ multiplet.
Both computations match very well and are in good agreement with the experiment as well (crosses at $e=0.05$). 
They also show that the first-order perturbation theory, given by the tangents at $e=0$, is not sufficient for ZoRo's ellipticity ($e=0.05$). This is clearly illustrated by the crossing between the $m=0$ and $|m|=1$ branches around $0.025$ (Fig.~\ref{fig:galcomsolvseps}).

\subsection{Second-order ellipticity perturbation}

A second-order perturbation theory for a homogeneous fluid enclosed in a quasi-spherical container has been developed  \cite{mehl1982acoustic, mehl1986acoustic, mehl2007acoustic}, calculating the eigenfrequencies of a slightly aspherical acoustic cavity resonator enclosed within a spherical cavity \cite{morse1953methods}.
Our implementation in a Matlab package of Mehl's method \cite{mehl2007acoustic} is described in Supplementary material\cite{supmatmain} with more details. We compared our results both with his original results and with finite-element calculations. 

To do so, we use the built-in modules of COMSOL Multiphysics, which solves the acoustic equation  for a given (arbitrary) shape. To single out the second-order correction and check the validity of the perturbation theory, we use Lagrange elements of degree $5$ to compute accurately the eigenfrequencies for a series of oblate spheroids with increasing ellipticity $0 < e < 10^{-3}$.
Then for each ${_n}\mathcal{S}_l^m$ mode, the evolution of its frequency with ellipticity is fit by a second-order polynomial as $ f_{nlm}(\epsilon) = f_{nl} + \gamma_{nlm}^{(1)} \epsilon + \gamma_{nlm}^{(2)} \epsilon^2$
with $\epsilon = e \, r_{eq}/r_{pol}$, in order to easily compare with Mehl's results \cite{mehl2007acoustic}. We have checked that the results are unchanged when higher orders are taken into account in this fit. 

\begin{figure}[ht]
\includegraphics[width=0.51\textwidth]{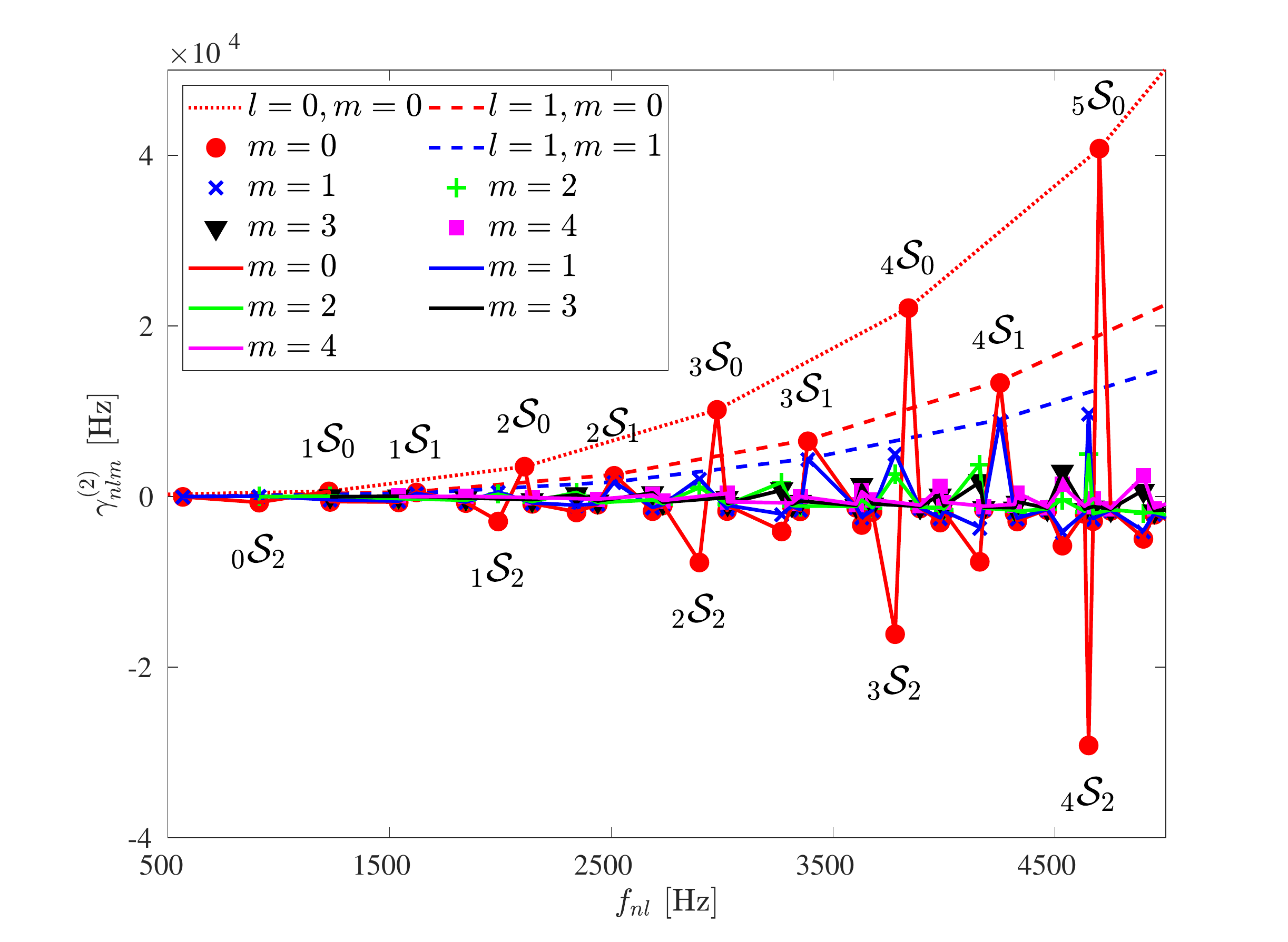}
\caption{\label{fig:gamma2}{Ellipticity second order $\gamma_{nlm}^{(2)}$ corrections for diffusionless acoustic eigenmodes in our experiment. We reproduced Mehl's original results \cite{mehl2007acoustic} for modes $l=0$ (connected by dotted line) and $l=1$ (connected by dashed lines) and compare them with our implementation of Mehl's theory \cite{mehl2007acoustic} (symbols) and with finite element calculations in oblate spheroids (dots connected by solid lines for a given $m$). All results from the three different methods are consistent.}}
\end{figure}

Figure~\ref{fig:gamma2} displays the evolution of $\gamma_{nlm}^{(2)}$ with $f_{nl}$ as computed with our various methods.
The solid (red) line joining the $m=0$ data displays a saw-tooth pattern, with the $l=1$ and $l=0$ points yielding increasingly large positive values, while the $l=2$ points get more and more negative as $f_{nl}$ increases.
In his article \cite{mehl2007acoustic}, Mehl only reports on the $l=0$ (dotted red line) and $l=1$ (dashed lines) cases.
The mode crossing we observe occurs for large enough negative value of $\gamma_{nlm}^{(2)}$ and was therefore not present in his results.
Our implementation of perturbation theory is in very good agreement with both Mehl's original results \cite{mehl2007acoustic} and the finite-element numerical calculations, showing that perturbation theory up to second-order is sufficient for mode identification in ZoRo within our frequency range.

\subsection{\label{subsec:spectracomparison} Comparison with experimental spectra}

Focusing on the ${_2}\mathcal{S}_2$ multiplet, the second-order correction for $m=0$ is negative while the other $m$ are positive (close to zero) (Fig.~\ref{fig:gamma2}), yielding \textit{in fine}  a lower frequency for ${_2}\mathcal{S}_2^0$ than ${_2}\mathcal{S}_2^1$.

\begin{center}
   \begin{figure}[htbp]
    \includegraphics[width=\textwidth]{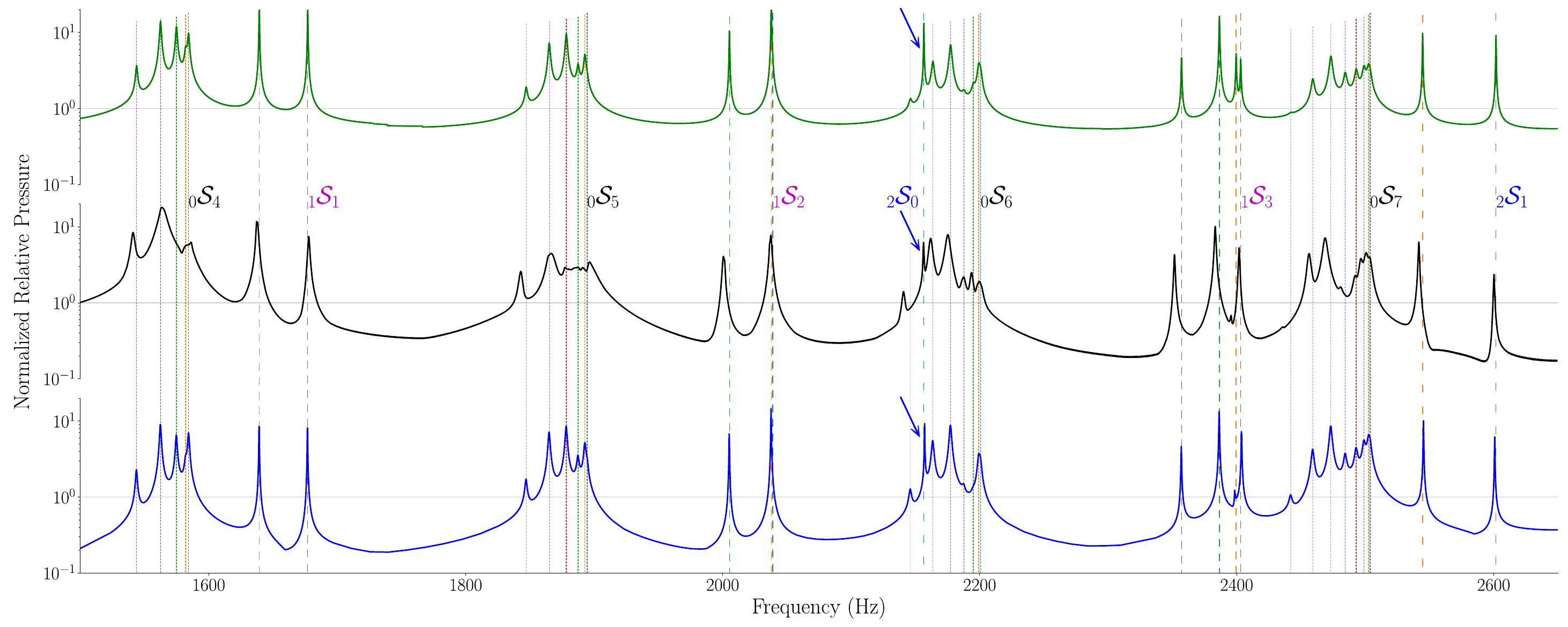}
    \caption{\label{fig:fullthexpcom}{Acoustic spectra for ZoRo configuration at rest obtained with perturbation theory (top), experimental data (middle) and finite-element calculations (bottom). Mode frequencies and labels from the theory are given across the three spectra for comparison (vertical lines, different line types are used for $n$, different colours for $m$). Experimental spectrum is averaged over all electrets of one hemisphere. For each spectrum, amplitude is normalised by its mean on the given frequency window.}}
    \end{figure}
\end{center}

In figure~\ref{fig:fullthexpcom} we plot the acoustic spectra from experimental data, perturbation theory and finite-element calculations.
The first row shows the synthetic spectrum obtained from the perturbation theory (with the second-order correction in geometry) for the sound speed $c=343.194$~m.s$^{-1}$, which is the typical value for dry air at $20^{\circ}$C. The second row shows the experimental data. For the whole experimental spectrum, we apply a multiplicative correcting factor on the frequency axis to align the frequency of the mode $_2\mathcal{S}_0^0$ (indicated by an arrow) with the one of the theory.
The third row displays finite-element computation of the spectrum.
For the sake of comparison with experiments and theory, we consider shear and bulk viscosities and no-slip and constant temperature boundary conditions, which prevents the use of the built-in acoustic equation of COMSOL. Instead, we have modified the built-in aero-acoustic interface, which is limited to $m=0$. This non-trivial extension to arbitrary $m$ is fully detailed in Supplementary material \cite{supmatmain}. In practice, we use Lagrange elements of order $3$ for the pressure and order $4$ for the velocity and temperature. We then obtain the acoustic response for each $m$, and the complete spectrum is obtained by summing contributions of all $m$ (Fig.~\ref{fig:fullthexpcom}, bottom). Note that a typical calculation of a unique fluid response takes $\sim 12$~s on a current desktop computer, and the accurate calculation of the high-resolution full spectrum on $500-3000$~Hz requires thus a full week.

Complete comparison of the predicted second-order frequencies with experimental spectrum shows that all major features are reproduced (peak frequencies differ by less than $1$~Hz).
Each experimental peak can be identified. 
Both frequencies (real part), relative amplitude of peaks and their shape (imaginary part) are very similar, except in some narrow frequency range (e.g. around 1600~Hz) where some experimental peaks seem to be missing.
The dissipation in the experiment is expected to be higher than from the values used in the theory, where we neglected air humidity and other diffusion mechanisms. This might cause some peaks to be hidden in the experimental spectrum by the neighbouring peaks (see Supplementary material\cite{supmatmain} for whole frequency range comparison).
Other minor discrepancies can be attributed to slight geometry perturbations due to the presence of the instrumentation (e.g. loudspeakers).
The numerical and theoretical spectra are very similar, showing that we can rely on the perturbative approach for an accurate and robust description of the experimental spectrum.
\onecolumngrid
\twocolumngrid

\section{\label{sec:Rot} Splittings due to solid-body rotation}

In this section, we consider solid-body rotating flows in ZoRo. To obtain these flows, we impose a constant rotation rate to the container and we wait for $2$ minutes, a long time compared to the spin-up time ($\sim 10$ s for air rotating at 10~Hz), such that the fluid is uniformly rotating with the container \citep{greenspan1968theory}. Then, the fluid is at rest in the frame rotating with the container.

\subsection{\label{subsec:exprot} Effects of the rotation on the $_0\mathcal{S}_2$ multiplet}
\begin{figure}[ht]
\includegraphics[width=0.49\textwidth]{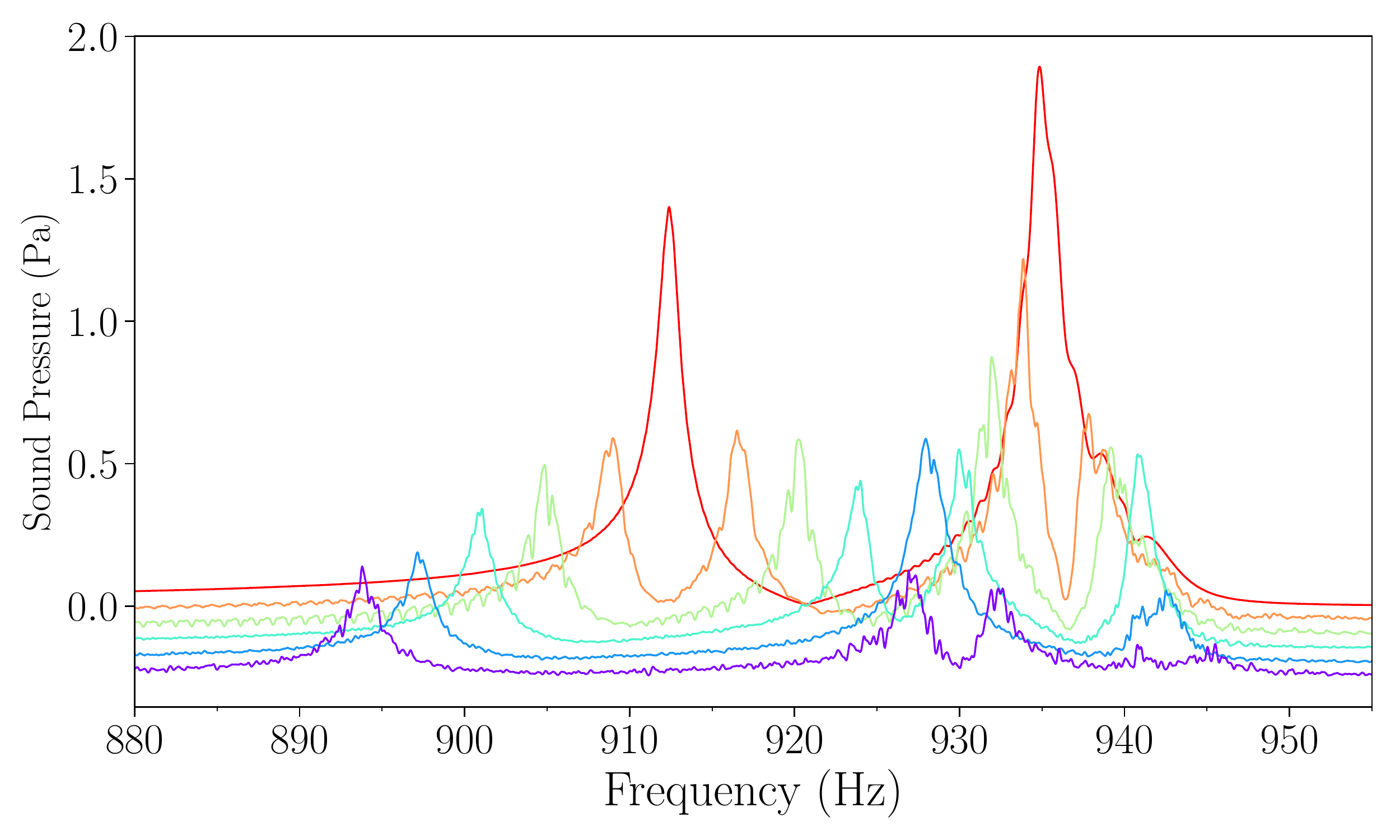}   
\includegraphics[width=0.49\textwidth]{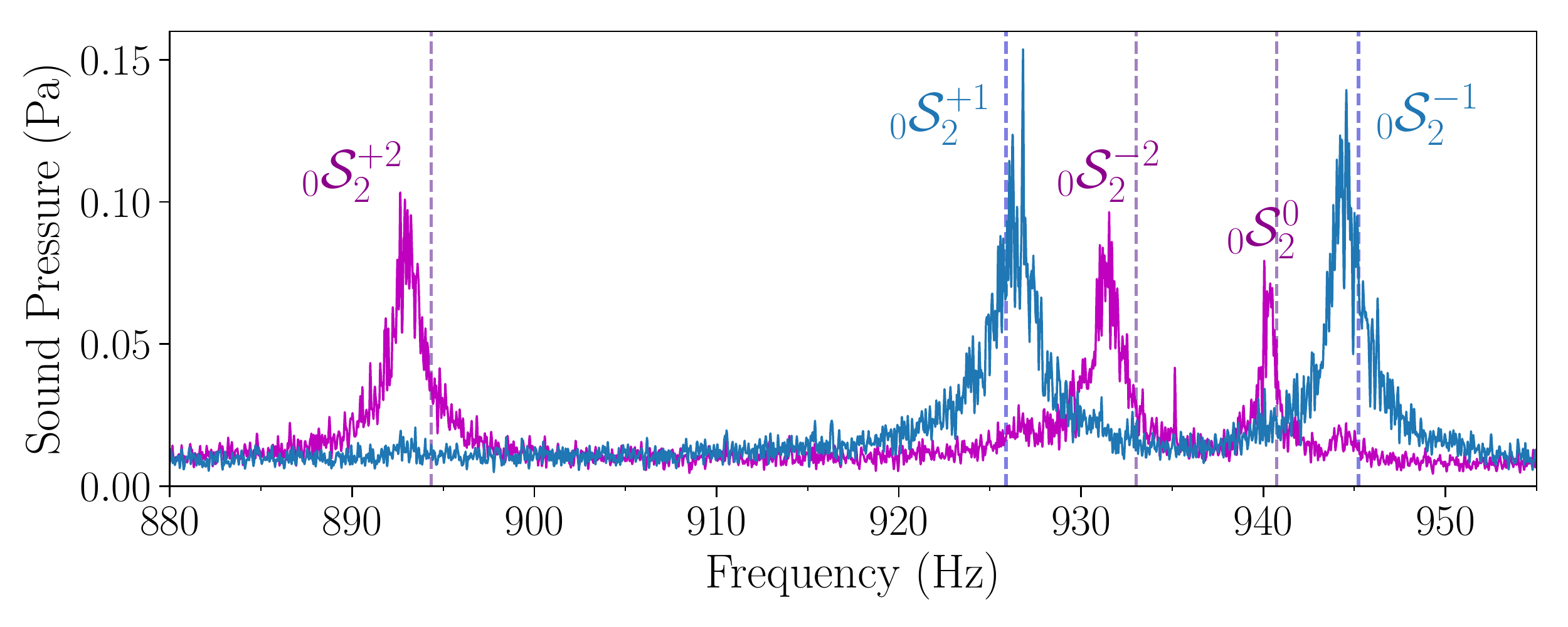}
\caption{\label{fig:sbrotexp}{Experimental spectra centered on $_0\mathcal{S}_2$. Top: for solid body rotation at increasing rotation rates from rest (top, red) to $f_{\Omega}=25$~Hz (bottom, violet) with $5$~Hz increment. Spectra are filtered (low-pass) and vertically shifted for visualisation purposes. Bottom: for solid body rotation at $f_{\Omega} =25$~Hz separated by symmetry, $m=\pm1$ (blue) and $m=0,\, \pm2$ (purple)  (colour online). The dashed lines shows the perturbation theory predictions.}}
\end{figure}

Using the protocol developed in Section~\ref{sec:Expspectra} to acquire and identify acoustic modes at rest, we measure the acoustic response in presence of rotation.
The top panel of figure~\ref{fig:sbrotexp} shows typical splittings due to solid-body rotations for the $_0\mathcal{S}_2$ multiplet. Focusing first on the original peak (i.e. with no rotation) around $910$~Hz, we observe that it splits into two peaks about half-height on both sides of the original peak.
We note that the splitting increases with the rotation rate.

In order to compare with the rotation splitting predicted by the theory, certain mode identification is needed, especially in regions where several peaks are close together, around 930~Hz for example. In the bottom panel of figure~\ref{fig:sbrotexp} we separate the $_0\mathcal{S}_2$ multiplet by symmetry at $f_\Omega =25$~Hz. We use the symmetry method detailed in Section \ref{subsec:modeid} to separate odd and even $m$ modes. We can also identify the $m=0$ mode around $940$~Hz by the fact that it is the only mode not influenced by rotation at first-order, as shown by equation (\ref{eq:deltaomROT}). 
The vertical lines show the frequencies given by the perturbative approach taking into account the first-order rotation effects (Section \ref{sec:Theory}). This allows to identify the mode $_0\mathcal{S}_2^{-2}$ at $930$~Hz and the mode $_0\mathcal{S}_2^{+1}$ at $926$~Hz: the two peaks cross each other. This lifts the ambiguity that could have arisen from a naive reading of the bottom purple curve ($f_{\Omega}=25$~Hz) in the top panel. At $f_{\Omega}=20$~Hz (blue curve), these $2$ peaks merge in a higher peak at $928$~Hz (see the top panel).

\subsection{\label{subsec:rottheory} First experimental determination of the Ledoux coefficients}
For a given ${_n}\mathcal{S}_l^m$ mode, perturbation theory predicts a linear increase of the rotational splitting with the rotation rate. 
To verify this prediction and its validity domain, we extract splittings for a collection of non-axisymmetric eigenmodes over our range of working rotation rates (from rest to $30$~Hz). 
We fit the observed $\pm m$ pairs of spectral peaks with synthetic spectra, carrying a grid search on the four following parameters: the frequency splitting between the two peaks, their mean frequency, their width, and their amplitude. 
For each combination of parameters, we evaluate the misfit as the root-mean square (rms) difference between the observed and synthetic spectra (in log scale) in the frequency window of the mode.
We obtain the best frequency splitting from the combination yielding the smallest misfit. The error is estimated from the minimum and maximum splittings for which some parameter combinations produce a misfit of typically $1.05$ times the minimum misfit.
We provide examples of the fits in Supplementary material\cite{supmatmain}.

In figure \ref{fig:splitrot}, we show a selection of experimental splittings as a function of rotation rate (the experimental splittings being measured by the difference between the $\pm m$ peak frequencies). At first glance, we observe that they agree well with the theoretical linear predictions $|2 \delta_{\Omega}|$ given by equation (\ref{eq:deltaomROT}) and shown by the lines. 
The blue crosses show our finite calculations predictions which are, in many cases, closer to the experimental data than the theory (see e.g. $_0\mathcal{S}_2^2$ or $_0\mathcal{S}_2^1$). In our range of rotation rates, the numerical splitting also follow a linear trend. Note however, that the observed slopes can be slightly different from the one predicted using Ledoux coefficients in the sphere (see typically $_0\mathcal{S}_2^1$). This is due to the effect of ellipticity \citep{vidal2019acoustics}, and this slightly different slope defines spheroidal Ledoux coefficients, which now depends on $\epsilon$ but also on $m$ by contrast with the spherical one (which is independent of $m$, see equation \ref{eq:Ledoux}).    

\begin{figure}[ht]
\includegraphics[trim=0 0 0 0, clip,width=1.1\reprintcolumnwidth]{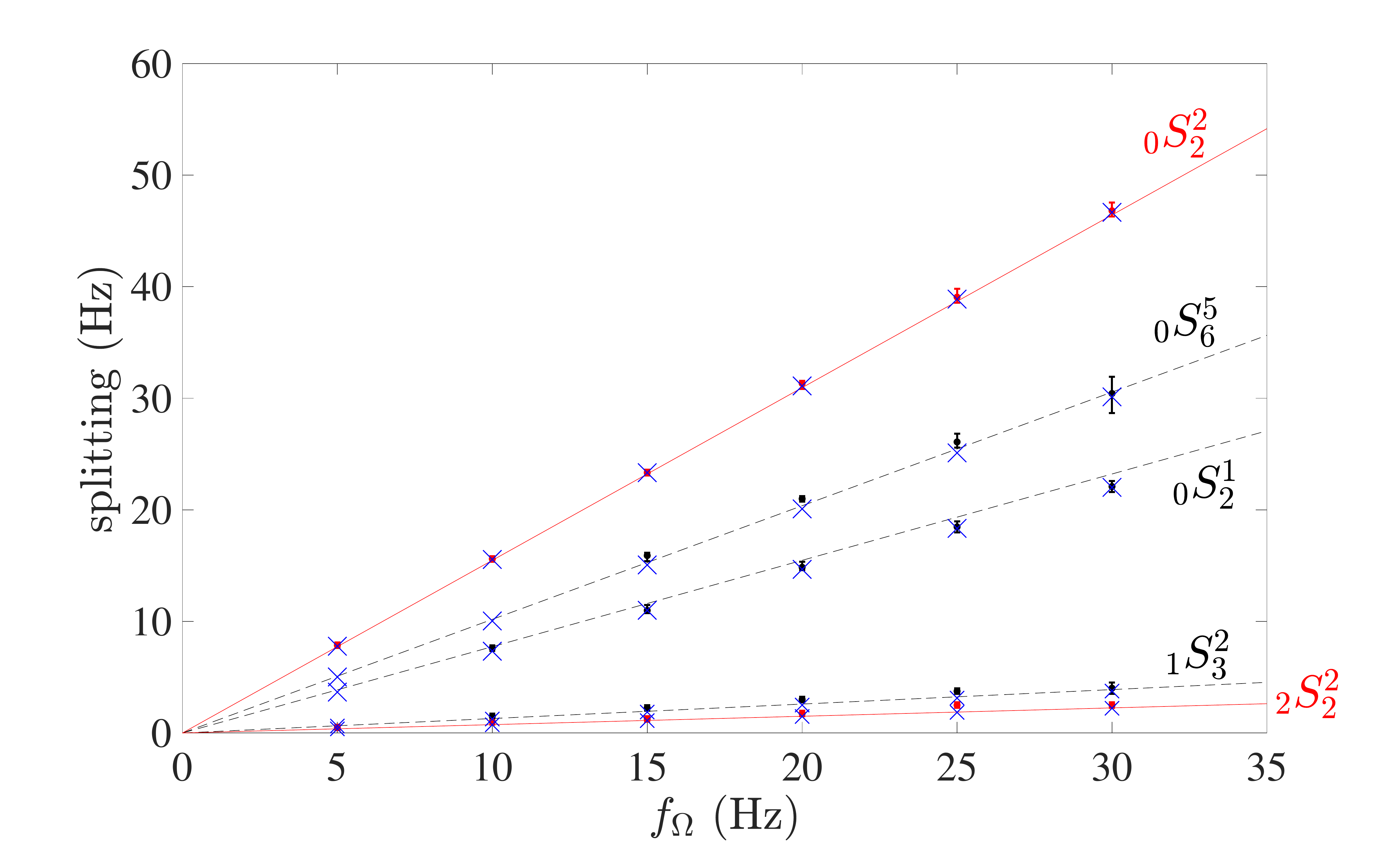}
\caption{Rotational splittings of chosen modes for increasing rotation rates. Experimental (circles) and finite element (blue crosses) splittings are measured by the difference between the $\pm m$ peak frequencies. Theoretical linear splittings (lines) are given by $|2 \delta_{\Omega}|$, see equation (\ref{eq:deltaomROT}). For theoretical splittings, equatorially symmetric modes are shown in red solid lines and anti-symmetric modes in black dashed line; we use the same choice of colours for the experimental splittings (colour online). 
}
\label{fig:splitrot}
\end{figure}
\begin{figure}[ht]
\includegraphics[width=0.52\textwidth]{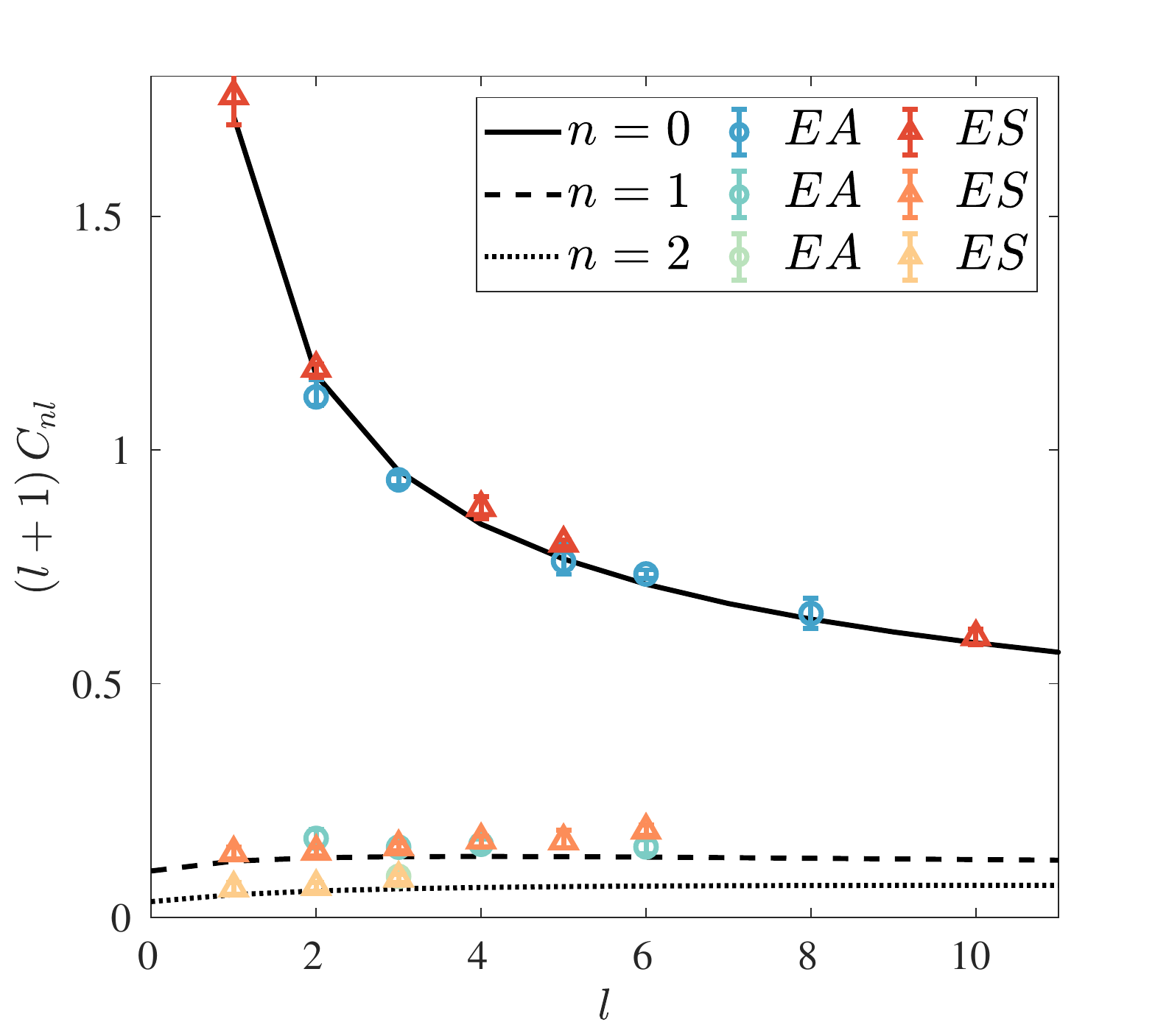}
\caption{\label{fig:ledoux}{Comparison between the theoretical Ledoux coefficients multiplied by $l+1$ (connected by lines for $n=0,\,1,\,2$) and their experimental counterparts deduced from the mode splitting measured in ZoRo (rotating at 20~Hz) for several $l$-modes of various $m$, being either symmetric ($ES$, hot-coloured triangles) or anti-symmetric ($EA$, cold-coloured circles) with respect to the equator (colour online). When possible, several $\pm m$ pairs are included for one $_n\mathcal{S}_l$ multiplet.}}
\end{figure}

In our range of rotation rates, rotational splittings in the sphere are predicted to be proportional to $f_\Omega$, but also to $m$ through the spherical Ledoux coefficients. 
By successively focusing on various non-axisymmetric multiplets, we retrieve the rotational splittings for a larger collection than before. 
We have extracted rotational splittings for 24 modes at $f_\Omega =20$~Hz with the method detailed above and computed the associated spheroidal Ledoux coefficients (frequency splitting divided by $2 m f_{\Omega}$). Figure \ref{fig:ledoux} shows these Ledoux coefficients for different $\pm m$ pairs of peaks as a function of $l$. For each mode, the experimental Ledoux coefficients (symbols) agree roughly with the linear theory (lines) and the deviations may be explained by the ellipticity of the container (see $l=2$, $n=0$ in figures \ref{fig:splitrot} and  \ref{fig:ledoux}). 
Conversely, the determination of the rotation rate can be obtained by measuring any splitting if the spheroidal Ledoux coefficient is known.

\section{\label{sec:Conclu} Conclusions and perspectives}
For this study, we built an experimental setup made of a gas-filled spheroid cavity (ZoRo) rotating up to $30$~Hz. We successfully model the experimental acoustic spectrum of the cavity with a perturbation theory and finite-element calculations. To identify the modes, we need to introduce a second-order geometry correction and use the equatorial symmetries of the modes, sources and receivers. We have successfully measured the Coriolis effects on the acoustic response, allowing the first experimental determination of the Ledoux coefficients \citep{ledoux1951nonradial} and their correction due to the ellipticity of the cavity\cite{vidal2019acoustics}.

We aim to use this modal acoustic velocimetry method to image flow fields in our ZoRo apparatus. Having identified modes up to $l=10$ and $n=3$, we can expect a spatial resolution of $1/5$ of the cavity radius \cite{triana2014}. This modal acoustic velocimetry is thus a robust, versatile and non-intrusive velocimetry method. As a long-term goal, we aim to study more complex large-scale rotating flows, including non-uniform rotation rates, and use acoustic splitting data to retrieve the time-dependent three components of flow velocity in the whole volume. It could also be interesting to increase rotation rates to probe the limits of the perturbation theory, to 
investigate higher-order Coriolis effects \cite{vidal2019acoustics} and centrifugal effects \cite{Ecotiere_Tahani_Bruneau_2004,reese2006acoustic}.
Another perspective is to use other internal gases with different diffusive behaviour, leading to a change of peaks width, potentially allowing better peak separation (SF$_6$ has thinner $n=0$ modes than air for example). 

We plan to use this modal acoustic velocimetry to describe the physics of zonal jets. Future planned experiments include changing (increasing or decreasing) the pressure within the spheroid and differential heating of the working gas to reproduce the forcings and forces balance seen in astro-geophysical bodies.

\begin{acknowledgments}
This research was supported by ANR-13-BS06-0010 (TuDy). JV was partly funded by STFC Grant ST/R00059X/1. HCN wishes to thank M. Moldover, M. Bruneau and C. Guianvarc'’h for their interest and encouragements.
This work is dedicated to the memory of J. Mehl, who gave the authors his very last encouragements to pursue the path he pioneered.
ISTerre is part of Labex OSUG@2020 (ANR10 LABX56).
The MATLAB package to calculate topographic and rotation influence on acoustic modes with perturbation theory is available at \url{https://www.isterre.fr/annuaire/member-web-pages/henri-claude-nataf/}. The COMSOL models are available at \url{https://www.isterre.fr/annuaire/member-web-pages/david-cebron/}. 
Most figures were produced using matplotlib (\url{http://matplotlib.org/}).  
\end{acknowledgments}

\onecolumngrid
\bibliography{biblio}
\bibliographystyle{prsa.bst}

\end{document}